\documentclass[conference]{IEEEtran}
\IEEEoverridecommandlockouts
\usepackage{cite}
\usepackage{amsmath,amssymb,amsfonts}
\usepackage{algorithmic}
\usepackage{graphicx}
\usepackage{textcomp}
\usepackage{xcolor}
\usepackage{todonotes}
\usepackage{xspace}
\usepackage{url}
\usepackage{hyperref}
\usepackage{flushend}
\def\BibTeX{{\rm B\kern-.05em{\sc i\kern-.025em b}\kern-.08em
    T\kern-.1667em\lower.7ex\hbox{E}\kern-.125emX}}

\newcommand{\sbft}{SBFT\xspace}

\newcommand{\rdp}{\textsc{Randoop}\xspace}
\newcommand{\evo}{\textsc{EvoSuite}\xspace}
\newcommand{\bbc}{\textsc{BBC}\xspace}
\newcommand{\evofz}{\textsc{EvoFuzz}\xspace}
\begin{document}

\title{\sbft Tool Competition 2025 - Java Test Case Generation Track}

\author{\IEEEauthorblockN{Fitsum Kifetew}
\IEEEauthorblockA{\textit{Software Engineering Unit} \\
\textit{Fondazione Burno Kessler}\\
Trento, Italy \\
kifetew@fbk.eu}
\and
\IEEEauthorblockN{Lin Yun}
\IEEEauthorblockA{\textit{Department of Computer Science and Engineering} \\
\textit{Shanghai Jiao Tong University}\\
Shanghai, China \\
lin\_yun@sjtu.edu.cn}
\and
\IEEEauthorblockN{Davide Prandi}
\IEEEauthorblockA{\textit{Software Engineering Unit} \\
\textit{Fondazione Burno Kessler}\\
Trento, Italy \\
prandi@fbk.eu}
}

\maketitle

\begin{abstract}
This short report presents the 2025 edition of the Java Unit Testing Competition in which four test generation tools (\evofz, \evo, \bbc, and \rdp) were benchmarked on a freshly selected set of 55 Java classes from six different open source projects. The benchmarking was based on structural metrics, such as code and mutation coverage of the classes under test, as well as on the readability of the generated test cases.  
\end{abstract}


\section{Introduction} \label{sec:introduction}

The Java unit test generation tool competition has been running for more than a decade, and this year we report on the thirteenth edition held at the \sbft 2025 workshop, co-located with ICST 2025. This year's edition evaluated four Java unit test generation tools:  \evofz~\cite{DBLP:conf/sbst/MoonJ24}, \evo~\cite{DBLP:journals/tosem/FraserA14}, \bbc~\cite{DBLP:journals/ese/DerakhshanfarDZ22}, and \rdp~\cite{DBLP:conf/oopsla/PachecoE07}. The evaluation involved  structural as well as readability metrics, on a benchmark composed of open source Java projects that were not used in previous editions of the competition. The idea of using non-structural metrics in the benchmarking process is similar to what has been done in recent editions of the competition~\cite{DBLP:conf/icse/JahangirovaT23}, however in this edition we assess the non-structural metric, test readability, by exploiting Large Language Models (LLMs).

All experiments were carried out on identical hardware configurations and using the benchmarking infrastructure~\cite{DBLP:journals/stvr/DevroeyGGJKPP23} which has been used in almost all previous editions of the competition.

The results of the benchmarking show that with respect to structural metrics, the \bbc tool is ranked first, while for readability metrics, \evo and \evofz have better performances. Overall the global ranking, considering both structural and readability metrics, the \bbc tool is ranked first.

In the rest of this report we briefly describe our benchmark in Section~\ref{sec:benchmark} followed by a short description of the tools in Section~\ref{sec:tools}. The methodology adopted is described in Section~\ref{sec:methodology} and results are presented in Section~\ref{sec:results}. Finally Section~\ref{sec:conclusion} concludes the report.

\section{The benchmark} \label{sec:benchmark}

\begin{table}[t]
\centering
\begingroup\fontsize{7pt}{8pt}\selectfont
\begin{tabular}{l|rr|rr|rr}
  \hline
   & \multicolumn{2}{|c|}{CUTs}  & \multicolumn{2}{|c|}{Sampled CUTs}  & \multicolumn{2}{c}{Final CUTs} \\
  Project & n. & CCN & n. & CCN & n. & CCN \\
  \hline
\emph{Thumbnailator} & 49 & 2.0 & 17 & 3.2 & 15 & 3.3 \\
  \emph{Querydsl} & 681 & 1.6 & 99 & 3.1 & 15 & 3.4 \\
  \emph{Java Faker} & 100 & 1.2 &  4 & 2.2 &  4 & 2.2 \\
  \emph{JavaCV} & 67 & 3.5 & 51 & 4.1 & 17 & 4.1 \\
  \emph{Easy Rules} & 56 & 1.7 &  6 & 4.0 &  3 & 2.2 \\
  \emph{Squash Java} &  2 & 2.4 &  1 & 2.4 &  1 & 2.4 \\
   \hline
  Total &  955 & - &  178 & - &  55 & - \\
   \hline
\end{tabular}
\endgroup
    \caption{Characteristics of the benchmark. The column 'CUTs' shows all the CUTs in the respective projects. The column 'Sampled CUTs' shows the CUTs we sampled based on CCN. The column 'Final CUTs' shows the final set of CUTs that are actually used for the ranking of the tools.}
    \label{tab:prjData}
\end{table}

The set of Java classes for this year's edition have been chosen from six open source projects spanning over different application domains, making sure that they were not used in previous editions of the competition. The six projects are:
\begin{itemize}
    \item \emph{Thumbnailator\footnote{\url{https://github.com/coobird/thumbnailator}}}: a thumbnail generation library for Java.
    \item \emph{Querydsl}\footnote{\url{https://github.com/querydsl/querydsl}}: a framework which enables the construction of type-safe SQL-like queries for multiple backends including JPA, MongoDB and SQL in Java.
    \item \emph{Java Faker}\footnote{\url{https://github.com/DiUS/java-faker}}: a library that generates fake data.
    \item \emph{JavaCV}\footnote{\url{https://github.com/bytedeco/javacv}}: a library that provides utility classes to make JavaCPP presets of computer vision libraries easily accessible from Java. 
    \item \emph{Easy Rules}\footnote{\url{https://github.com/j-easy/easy-rules}}: a Java rules engine.
    \item \emph{Squash Java}\footnote{\url{https://github.com/airbnb/squash-java}}: a Squash client library that exports exception to Squash.
\end{itemize}

To make sure that the classes pose a reasonable challenge to the tools in the competition, we have filtered them based on their cyclomatic complexity (CCN) and discarded classes with low CCN. Table~\ref{tab:prjData} presents a summary of the classes in the benchmark where we can see the total number of classes in all the projects (955 classes) and the resulting set of classes after filtering based on CCN (178 classes). As reported later in Section~\ref{sec:methodology}, several executions of the tools on the sampled classes resulted in errors. Hence, a large number of executions were excluded from the computation of the final ranking, resulting in a reduced number of classes (55) on which the final ranking of the tools was actually based. Note that the project CCN values reported in Table~\ref{tab:prjData} are computed as the average CCN of the classes in the project, and the CCN of a class is computed as the average CCN of its methods.

\section{The tools} \label{sec:tools}

This year's edition of the competition involves four tools: \evo, \evofz, \bbc, and \rdp.

\evo~\cite{DBLP:journals/tosem/FraserA14} is a tool for the generation of unit tests for Java programs. \evo contains a wide variety of algorithms for the test generation, including several evolutionary algorithms, which the user can select at runtime. The default algorithm, as used in this edition, is the DynaMOSA algorithm~\cite{DBLP:journals/tse/PanichellaKT18}.

\evofz~\cite{DBLP:conf/sbst/MoonJ24} is a tool based on \evo that employs fuzzing during test generation for rendering the process more effective. \evofz uses the DynaMOSA algorithm.

\bbc~\cite{DBLP:journals/ese/DerakhshanfarDZ22} is a tool based on \evo which introduces a new secondary objective called Basic Block Coverage, hence the name \bbc. This secondary objective is used to distinguish two individuals (test cases) that have the same fitness value by computing their respective coverage of the basic blocks between the nearest covered control dependent basic block and the target statement.

\rdp~\cite{DBLP:conf/oopsla/PachecoE07} is a tool that generates unit tests by employing feedback-directed random testing. Throughout the years, \rdp was used as a baseline tool in every edition of the tool competition.

It is worth noting that \evofz, and \bbc share many aspects with \evo, including the use of the DynaMOSA algorithm. Their difference is mainly in how the evolutionary search is performed where \bbc introduces a secondary objective corresponding to basic block coverage while \evofz introduces strategic fuzzing of primitive values in generated test cases.

\section{Methodology} \label{sec:methodology}

The methodology adopted for this year's edition of the competition is similar to previous years~\cite{DBLP:conf/icse/JahangirovaT23}, using the same benchmarking framework~\cite{DBLP:journals/stvr/DevroeyGGJKPP23}. Continuing with the tradition of the recent years' editions, also this year the evaluation criterion is composed of two aspects: structural metrics and readability metrics. The structural metrics are the traditional metrics used in automated testing, i.e., line, branch, and mutant coverage. On the other hand, the readability metric is meant to assess how readable the automatically generated tests are for humans.

\subsection{Structural coverage metrics} \label{sec:structural-metrics}
While it is customary to use multiple time budget settings and replications for the benchmarking, in this year's edition, due to technical problems we were able to use one time budget value of 120 seconds. Each tool is executed on each CUT three times, to account for the inherent randomness of the test generation process. Overall, given the  55 CUT, four tools, and three repetitions there were a total of 55 $\times$ 4 $\times$ 3 = 660 executions that were used to compute the final ranking of the tools.

It is important to note here that, the actual number of experiments performed is much larger, this is because that executing the competing tools on the initial set of CUTs selected for this year's benchmark resulted in several cases of failure that did not result in tests being generated. Hence, these data points were excluded from the rank computation. To give some context, the initial set of CUTs we selected were 178 spanning six projects (see Table~\ref{tab:prjData}).

The benchmarking was performed on machines with AMD EPYC 7302 CPUs with  speed of 3 GHz and 32 GB of RAM ensuring that all tools were executed in identical settings.

To compute the structural metrics, for each tool we calculated line, branch, and mutation coverage as in previous editions~\cite{DBLP:conf/icse/JahangirovaT23}. Similarly the same statistical analyses were carried out as in previous years to compare the different tools and to check for statistical significance~\cite{DBLP:conf/sbst/GambiJRZ22}.

\subsection{Readability metrics} \label{sec:readability-metrics}
To assess the readability of the generated test cases, one would perform a comprehensive study involving human evaluators that read the test cases and adjudicate them as more or less readable on a suitably established scale~\cite{DBLP:conf/icse/JahangirovaT23}. However, given the time and resource constraints typically associated with the testing tool competitions, we opted for another alternative that exploits the capability of LLMs in this regard. In particular, LLMs have shown impressive performance in understanding both natural language and code. Hence in this year's edition, we explored the use of LLMs for assigning readability scores to automatically generated test cases. We performed an exploratory study to verify whether LLMs could indeed be used for this purpose. The results we obtained were positive in that the LLM could give readability scores to a given test case and provide explanations for the scores, which we found to be plausible. Furthermore, we also observed that the response was consistent over different prompts. Subsequently, we developed a small Python program that parses a given Java test suite and prompts an LLM to get the readability scores. 

To assess readability, we considered four dimensions: clarity~\cite{winkler2024investigating}, naming conventions~\cite{oliveira2020evaluating}, simplicity~\cite{buse2008metric}, and structure~\cite{winkler2024investigating}. In particular, we query the LLM with the following  prompt:

\noindent \texttt{Please evaluate the readability of the following test case. Consider the following criteria:} \\
\\
\noindent \texttt{Clarity: Are the intentions and logic of the test easy to understand?} \\
\noindent \texttt{Naming Conventions: Are names meaningful and self-explanatory?} \\
\noindent \texttt{Simplicity: Is the code free from unnecessary complexity or redundant steps?} \\
\noindent \texttt{Structure: Is the test modular, organized, and easy to follow?} \\

The prompt is completed with the code of the test case to evaluate and instructions about the output format. In particular, we asked for a \emph{score out of 100} and \emph{less than 10 words explanation}.

Given that the four tools generated thousands of test cases, we sampled test cases in such a way that for each CUT we randomly selected three test cases for each tool. We then computed the readability scores for each test case using our Python program as described above and using Google$^\copyright$ Gemini 1.5 Flash~\cite{team2024gemini}.

\section{Results} \label{sec:results}

\begin{table}[htb]
\centering
\begingroup\fontsize{7pt}{8pt}\selectfont
\begin{tabular}{lrrr}
  \hline
Tool & Total n. tests & Mean n. tests & Gneration time \\
  \hline
Randoop & 1,561,533 & 9521.5 & 5h 35m \\
  EvoSuite & 2,049 & 13.0 & 2h 51m \\
  EvoFuzz & 4,535 & 28.2 & 4h 10m \\
  BBC & 4,238 & 27.5 & 4h 3m \\
   \hline
\end{tabular}
\endgroup
    \caption{Tests generation summary data}
    \label{tab:testsSummary}
\end{table}

Table~\ref{tab:testsSummary} provides a summary of the data regarding the number of tests generated and the time required for generation by each tool. \rdp generates a huge number of tests, producing more than three hundred times the amount generated by the other tools. At the other end, \evo creates half of the tests generated by \evofz and \bbc. This could be due to \evo's minimization process, as previously noted\cite{DBLP:conf/icse/JahangirovaT23}.
Looking at the generation time, \evofz and \bbc improve with respect to \rdp, but they require an hour more than \evo. Detailed experimental data for each tool, including analysis scripts, can be found in the online replication package~\cite{replication_package}.

\subsection{Structural coverage metrics}

\begin{figure}[htb]
    \includegraphics[scale=1]{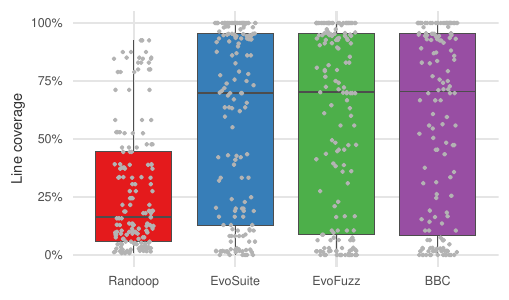}
    \caption{Line Coverage}
    \label{fig:lineCoverage}
\end{figure}

\textbf{Line Coverage}
Figure~\ref{fig:lineCoverage} shows the line coverage achieved across all CUTs. \rdp has the lowest median value (16.3\%), while the other 3 tools have comparable values ($\sim$70\%).
Conover's all-pairs rank comparison test showed no statistically significant difference among \evo, \evofz, and \bbc.

\begin{figure}[htb]
    \includegraphics[scale=1]{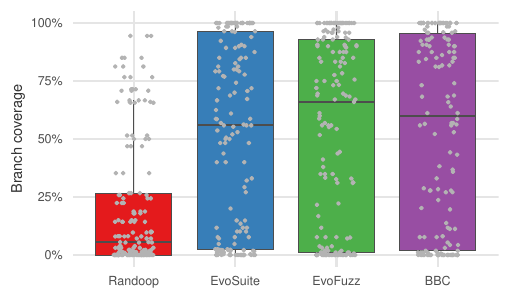}
    \caption{Branch Coverage}
    \label{fig:branchCoverage}
\end{figure}

\textbf{Branch Coverage}
When evaluating branch coverage, the ranking of the tools remains consistent, as illustrated in Figure~\ref{fig:branchCoverage}. The median coverage for \rdp stands at 5.6\%, which is significantly lower than that of \evo at 56.1\%, \bbc at 59.9\%, and \evofz at 66\%. However, the differences among the last three tools are not statistically significant.

\begin{figure}[htb]
    \includegraphics[scale=1]{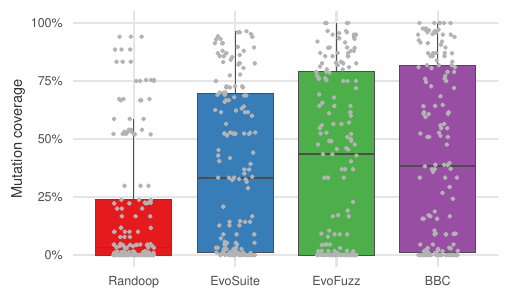}
    \caption{Mutation Coverage}
    \label{fig:mutationCoverage}
\end{figure}

\textbf{Mutation Coverage}
Mutation testing proved to be more complex with a significant number of cases where the coverage is zero. \rdp kills zero mutants in 38\% of the cases. Even if the other tools lower this percentage, the observed zero values are about 20\%. Figure~\ref{fig:mutationCoverage} shows the complete mutation coverage data. \evo's median mutation coverage (33\%) is lower than that of \bbc (38\%) and \evofz (43\%). However, these differences are not significant, according to Conover's all-pairs rank comparison test.

\subsection{Readability}
\begin{figure}[htb]
    \includegraphics[scale=1]{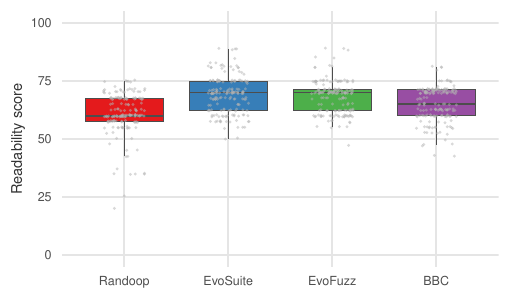}
    \caption{Readability Score}
    \label{fig:readability}
\end{figure}

Figure~\ref{fig:readability} shows the overall readability score. \evo and \evofz have the same median score (70), while \bbc shows a lower score (65), followed by \rdp (60). Interestingly, the differences between \bbc and the top performing tools are significant according to Conover's all-pairs rank comparison test. On a closer look, \bbc and \evo/\evofz have comparable scores when considering clarity and naming conventions, while simplicity and  structure are significantly lower in \bbc.

\subsection{Overall Results}
Table~\ref{tab:finalRankings} presents the overall ranking achieved by the considered tools. When considering the structural coverage criteria, \bbc has the nest ranking, followed by \evofz, \evo, and \rdp.
Note that the first three tools did not show statistically significant differences in any of the structural coverage criteria considered. Considering the readability metric, \evo has the best rank, followed by \evofz, \bbc, and \rdp. As we already noticed, the difference between \evo and \evofz is not statistically significant. To compute the final score, taking into consideration the fact that using LLMs to estimate readability is quite experimental, we limited the contribution of the readability score to 10\% of the overall score, as announced in the call for contributions\footnote{\url{https://sbft25.github.io/tools/java}}. The final ordering of the considered tools is \bbc, \evofz, \evo, and \rdp.

\begin{table}[htb]
\centering
\begingroup\fontsize{7pt}{8pt}\selectfont
\begin{tabular}{lrrr}
  \hline
     & \multicolumn{3}{c}{Ranking} \\
Tool & Coverage & Readability  & Overall  \\
  \hline
BBC & 2.03 & 2.68 & 2.09 \\
  EvoFuzz & 2.15 & 2.05 & 2.14 \\
  EvoSuite & 2.55 & 1.94 & 2.48 \\
  Randoop & 3.27 & 3.34 & 3.28 \\
   \hline
\end{tabular}
\endgroup
\caption{Final Rankings}
\label{tab:finalRankings}
\end{table}

\section{Conclusion} \label{sec:conclusion}

This year's edition featured four tools and a benchmark with a large number of classes, however during the execution phase we discovered that several of the classes caused different runtime errors across all the tools, hence we were forced to exclude them, reducing the benchmark size to 55 classes. While these failures merit further investigation to exclude potential problems within the benchmarking infrastructure, the final set of 55 classes is a reasonable number for assessing the four tools. Overall, the results did not show a marked difference among the four tools, except for \rdp. This could be due to the fact that \evofz and \bbc are based on \evo and hence have many similarities. We also observe that the structural metrics results are overall lower than what is typically observed previously~\cite{DBLP:conf/icse/JahangirovaT23}. This could be of interest to the respective tool developers as this year's benchmarks represent a set of classes that pose greater difficulty to the automated test generation tools.

\section*{Acknowledgment}
We would like to thank Shaker Mahmud Khandaker (from Fondazione Bruno Kessler) and Yuhuan Huang (from Shanghai Jiao Tong University) for their contributions in implementing the readability metric, benchmark selection, and running of the experiments.

\bibliographystyle{IEEEtran}
\bibliography{references}

\end{document}